\documentclass[12pt]{iopart}
\usepackage{graphicx}
\usepackage{bm}
\newcommand{\beq}{\begin{equation}}
\newcommand{\eeq}{\end{equation}}
\newcommand{\beqa}{\begin{eqnarray}}
\newcommand{\eeqa}{\end{eqnarray}}
\def\ra{\rangle}
\def\la{\langle}

\newcommand {\fexp} [1] {\exp \left[ #1 \right]}

\begin{document}
\title{Transitionless quantum drivings for the harmonic oscillator}

\author{J. G. Muga$^{1}$, X. Chen$^{1,2}$, S. Ib\'a\~nez$^{1}$, I. Lizuain$^{1}$, A. Ruschhaupt$^{3}$  
}

\address{$^{1}$ Departamento de Qu\'{\i}mica-F\'{\i}sica,
UPV-EHU, Apdo 644, 48080 Bilbao, Spain}

\address{$^{2}$ Department of Physics, Shanghai University,
200444 Shanghai, P. R. China}

\address{$^{3}$ Institut f\"ur Theoretische Physik, Leibniz
Universit\"{a}t Hannover, Appelstra$\beta$e 2, 30167 Hannover,
Germany}

 


\begin{abstract}
Two methods to change a quantum harmonic oscillator frequency
without transitions in a finite time are described and compared.
The first method, a transitionless-tracking algorithm,
makes use of a generalized 
harmonic oscillator and a non-local potential. The second method,
based on engineering an invariant of motion, only modifies the harmonic 
frequency in time, keeping the potential local at all times.      
\end{abstract}

\section{Introduction}
Changing the external parameters of the Hamiltonian 
is a fundamental and standard operation to probe, control, or prepare
a quantum system. 
In many cases it is desirable to go from an   
initial parameter configuration to a final one without
inducing transitions, as in the expansions performed in 
fountain clocks \cite{Bize}. In fact most
of the current experiments with cold atoms are based on a cooling stage 
and then an adiabatic drive of the system to some desired final trap or regime \cite{Polkov}. These ``transitionless'' \cite{Berry09}, or ``frictionless'' \cite{Ronnie} adiabatic processes may require exceedingly large times and become impractical, even impossible \cite{Polkov},  or quite simply a faster process is desirable, e.g. to increase
the repetition rate of a cycle, or a signal-to-noise ratio. 
This motivates the generic objective of 
achieving the same final state as the slow adiabatic processes, 
possibly up to phase factors, 
but in a much shorter time. One may try to fulfill that goal  
in two different ways: (a) designing appropriate ``parameter
trajectories'' of the Hamiltonian from the initial to the final times,
or (b) applying entirely new interactions that modify the Hamiltonian  
beyond a simple parameter evolution of the original form, for example by adding 
different terms to it. In this paper we shall analyze  
and discuss, for the harmonic oscillator,
two recently proposed methods whose relation had not been investigated. 
It turns out that they actually implement these two different routes.
While most of the treatment is applicable to an ``abstract'' harmonic oscillator, 
we shall discuss physical implementations specific of ultracold atoms or ions. Indeed, harmonic traps and their manipulation are basic working horses of this field.     

For the harmonic oscillator the parameter we consider is the trap frequency, 
which should go from    
$\omega_0$ to $\omega_f$
in a time $t_f$, preserving the populations of the levels, $P_n(t_f)=P_n(0)$.  ``$n$'' labels the instantaneous $n$-th eigenstate of the initial and final harmonic oscillator Hamiltonians,
\beqa
H_0(0)|n(0)\ra&=&\hbar\omega_0(n+1/2)|n(0)\ra,
\nonumber\\
H_0(t_f)|n({t_f})\ra&=&\hbar\omega_f(n+1/2)|n({t_f})\ra.
\label{hot}
\eeqa

One of the methods we shall discuss here 
relies on a general framework set by Kato in a proof of the  adiabatic theorem \cite{Kato}, and has been formulated recently by Berry \cite{Berry09}. We shall term it ``transitionless-tracking''
approach, or TT for short; the other one \cite{harmo,becs} engineers the Lewis-Riesenfeld invariant \cite{LR69} by an inverse method \cite{Palao} to satisfy the desired boundary conditions; we shall call this method ``inverse-invariant'', or II for short.
In the basic version of TT the dynamics is set to follow at all 
intermediate times the adiabatic path defined by an auxiliary Hamiltonian $H_0(t)$ (in our case a regular harmonic oscillator with frequency $\omega(t)$ and boundary conditions $\omega(0)=\omega_0$ and $\omega_f=\omega(t=t_f)$),  and its instantaneous eigenvectors $|n(t)\ra$, up to phase factors. Instead, in the II approach the auxiliary object is an engineered Lewis-Riesenfeld invariant $I(t)$ set to commute with $H_0(0)$ at $t=0$ and with $H_0(t_f)$ at $t_f$. In both cases  
intermediate states may be highly non-adiabatic with respect to the instantaneous eigenstates of the Hamiltonians actually applied, $H_{TT}(t)$ and $H_{II}(t)$.     
 
We shall provide first the equations characterizing the two approaches
and then comment on possible physical implementations.
\section{Transitionless tracking algorithm}
\subsection{General formalism}
For the general formalism we follow \cite{Berry09} closely. 
Assume a time-dependent Hamiltonian $H_0(t)$ with initial and final 
values (\ref{hot}),  
instantaneous eigenvectors
$|n(t)\ra$ and eigenvalues $E_n(t)$,  
\beq
H_0(t)|n(t)\ra = E_n(t)|n(t)\ra.
\eeq
A slow change would preserve the eigenvalue and eigenvector along the  
dynamical evolution times a phase factor,
\beq
|\psi_n(t)\ra = \exp\left\{
-\frac{i}{\hbar}\int_0^{t} dt' E_n(t') -\int_0^t dt' \la n(t')|\partial_{t'}n(t')\ra\right\}|n(t)\ra.
\label{22}
\eeq
We now seek a Hamiltonian $H(t)$ such that the adiabatic approximation
$|\psi_n(t)\ra$ represents the {\it exact} dynamics, 
\beq
i\hbar\partial_t|\psi_n(t)\ra=H(t)|\psi_n(t)\ra. 
\eeq
$H(t)$ (which is $H_{TT}$ if distinction with the other method is needed)
is related to the corresponding unitary operator by
\beqa
i\hbar\partial_t U(t)&=&H(t)U(t),
\\
\label{ht}
H(t)&=&i\hbar(\partial_t U(t))U^\dagger(t).
\eeqa
Choosing 
\beq
U(t)=\sum_n\exp\left\{-\frac{i}{\hbar}\int_0^t dt' E_n(t')- \int_0^t dt'\la n(t')|\partial_{t'} n(t')\ra\right\}
|n(t)\ra\la n(0)|, 
\eeq
we find from (\ref{ht}),
\beq
\hat{H}(t)=\sum_n|n\ra E_n\la n|
+i\hbar\sum_n(|\partial_t n\ra\la n|-\la n|\partial_t n\ra|n\ra\la n|)
\equiv \hat{H}_0+\hat{H}_1,    
\eeq
where we have simplified the notation, $|n\ra=|n(t)\ra$.  
It is also possible to choose other phases in (\ref{22}) \cite{Berry09}. 
The simplest case is   
$U(t)=\sum |n(t)\ra\la n(0)|$, without phase factors, which leads to  
$H(t)=i\hbar\sum |\partial_t n\ra\la n|$.  
Note that with this choice 
$H_0(t)$ has been formally suppressed in $H(t)$ but still plays a role through
its eigenfunctions $|n(t)\ra$.  
\subsection{Application to the harmonic oscillator}
We now apply the above to the harmonic oscillator 
\beq
\hat{H}_0(t)=\hat{p}^2/2m+\omega(t)^2 \hat{x}^2/2m
=\hbar\omega(t)(\hat{a}^\dagger_t \hat{a}_t +1/2),
\label{ho}
\eeq
where $\hat{a}_t$ and $\hat{a}_t^+$ are the 
(Schr\"odinger picture!) annihilation and creation operators at time $t$,    
\beqa
\hat{x}&=&\sqrt{\frac{\hbar}{2m\omega(t)}}(a_t^\dagger +a_t),
\\
\hat{p}&=&i\sqrt{\frac{\hbar m\omega(t)}{2}}(a_t^\dagger -a_t),
\\
\hat{a}_t&=&\sqrt{\frac{m\omega(t)}{2\hbar}}
\left(\hat{x}+\frac{i}{m\omega(t)}\hat{p}\right),
\\
\hat{a}^\dagger_t&=&\sqrt{\frac{m\omega(t)}{2\hbar}}
\left(\hat{x}-\frac{i}{m\omega(t)}\hat{p}\right).
\eeqa
This time dependence may be misleading and a bit unusual at first so
we insist: since the frequency depends on time 
the ``instantaneous'' ladder operators $\hat{a}_t$, $\hat{a}^\dagger_t$
create or annihilate different 
``instantaneous'' states, adapted to the corresponding frequency. Thus, ladder operators with different time labels do not commute in general, although some combinations, e.g. those equivalent to powers of $\hat{x}$ and/or $\hat{p}$, 
do commute, as we shall see later.  

The instantaneous eigenstates $|n(t)\ra$ can be written
in coordinate representation as 
\beq
\la x|n(t)\ra =\frac{1}{\sqrt{2^n n!}}\left(\frac{m \omega (t)}{\pi \hbar}\right)^{1/4}\exp{\left(-\frac{1}{2}\frac{m \omega (t)}{\hbar} x^2\right)}H_{n}\left(\sqrt{\frac{m \omega (t)}{\hbar}}x\right), 
\eeq
and their derivative 
with respect to $t$ is 
%
%
\beqa
\la x |\partial_t n(t) \ra = \left(\frac{1}{4} - \frac{m \omega (t)}{2 \hbar} x^2 \right)\frac{\dot{\omega}}{\omega(t)} |n \ra
+ \sqrt{\frac{m \omega (t)}{2 \hbar}}x  \frac{\dot{\omega}}{\omega(t)} \sqrt{n} | n-1 \ra.
\eeqa
We find, using the recursion relation of Hermite polynomials
and their orthogonality,
\beqa
\la k|\partial_t n \ra= \left\{
\begin{array}{ll}
\frac{1}{4}\sqrt{n(n-1)}\frac{\dot{\omega}}{\omega(t)} &~~~ k=n-2 \\ \\
-\frac{1}{4}\sqrt{(n+1)(n+2)}\frac{\dot{\omega}}{\omega(t)}    &~~~ k=n+2 \\ \\
0  &~~~ (\mbox{otherwise})
\end{array}
\right.,
\eeqa
%
so that $\hat{H}_1 (t)$ can be written as
\beqa
\hat{H}_1 (t)&=& i \hbar \sum_n |\partial_t n \ra \la n| \equiv i \hbar \frac{\dot{\omega}}{\omega(t)}\sum_n \Bigg[\left(\frac{1}{4} - \frac{m \omega (t)}{2 \hbar} \hat{x}^2 \right)|n \ra \la n|
\nonumber
\\
&+& \sqrt{\frac{m \omega (t)}{2 \hbar}}\hat{x} \sqrt{n} | n-1 \ra \la n|\Bigg].
\eeqa
%
%
%
%
%
Using $a_t=\sum_n  \sqrt{n} |n-1(t) \ra\la n(t)|$, and the relations between 
$\hat{x}$, $\hat{p}$, $\hat{a}_t$ and $\hat{a}_t^\dagger$ written above, 
\beqa
\hat{H}_1 (t)&=& i \hbar \frac{\dot{\omega}}{\omega(t)}\sum_n \left[\frac{1}{4} - \frac{m \omega (t)}{2 \hbar} \hat{x}^2  
+ \sqrt{\frac{m \omega (t)}{2 \hbar}}\hat{x}   \hat{a}_t \right]
\nonumber
\\
%
%
&=&\frac{i \hbar}{4}\frac{\dot{\omega}}{\omega(t)} - \frac{1}{2}\frac{\dot{\omega}}{\omega(t)} \hat{x} \hat{p}. 
\eeqa
Using $[\hat{x}, \hat{p}]= i \hbar$, we finally write the Hamiltonian $\hat{H}_1(t)$
in the following simple forms 
\beq
\hat{H}_1(t)=-\frac{\dot{\omega}}{4\omega}(\hat{x}\hat{p}+\hat{p}\hat{x})
=i\hbar\frac{\dot{\omega}}{4\omega}[{\hat{a}}^2-({\hat{a}}^\dagger)^2].
\label{h1a}
\eeq
In the last expression the subscript $t$ in $\hat{a}$ and $\hat{a}^\dagger$ has been dropped 
because the squeezing combination ${\hat{a}}^2-(\hat{a}^\dagger)^2$ is actually independent
of time, so one may evaluate it at any convenient time, e.g. at $t=0$.
The connection with squeezing operators is worked out in the appendix. 

$H_1$ is therefore a non-local operator,
and does not have the form of a regular  harmonic oscillator 
potential with an $x^2$ term.
Nevertheless the final Hamiltonian $H=H_0+H_1$ is still quadratic in 
$\hat{x}$ and $\hat{p}$, so it may  be considered a generalised harmonic oscillator \cite{gho}.   
\subsection{Physical realization}
The nonlocality of $\hat{H}_1$, with a constant prefactor,  
can be realized in a laboratory 
by means of 2-photon Raman transitions for trapped ions
\cite{Itano,Zeng}. 
Since we have to evaluate as well 
the possibility of making the prefactor in $\hat{H}_1$ time dependent we need to 
provide the derivation with some detail, first for a time-independent $\omega$. 
\subsubsection{Raman two-photon transition in a trapped ion}
\begin{figure}[t]
\begin{center}
\includegraphics[width=5.5cm]{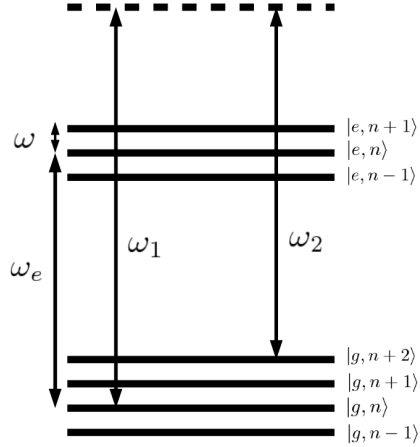}
\caption{Schematic electronic and vibrational level structure for a two-photon transition in an ion
trapped with frequency $\omega$. $\omega_1$ and $\omega_2$ are the laser
frequencies, and $\omega_e$ the transition frequency between ground and excited states. See the text for further details.}
\label{level_scheme_fig}
\end{center}
\end{figure}

Let us consider a harmonically trapped two-level system in 1D driven by two different lasers (with coupling strengths $\Omega_j$ and frequencies $\omega_j$, $j=1,2$), see Fig. \ref{level_scheme_fig} and Refs. \cite{zeng95,zeng98}. The time dependent ``Raman'' Hamiltonian in the Schr\"odinger picture will be given by
\begin{eqnarray}
\hat{H}_{R}(t)&=&\hat{H}_T+\hat{H}_A+\hat{H}_{int},
\end{eqnarray}
with ``trap'' ($T$), ``atomic'' ($A$), and interaction ($int$) terms
\begin{eqnarray}
\hat{H}_T&=&\hbar\omega \hat{a}^\dag \hat{a},\\
\hat{H}_A&=&\hbar\omega_e|e\ra\la e|,\\
\hat{H}_{int}&=&\sum_{j=1}^2\hbar\Omega_j\cos\left(\omega_jt-k_jx+\phi_j\right) (|g\ra\la e|+|e\ra\la g|),
\end{eqnarray}
where $\hbar\omega_e$ is the energy of the excited state $|e\ra$ and ${\bf k}_j=k_j{\bf \hat x}$ the wavevector of each laser which are assumed to be pointing along the principal trap direction, the $x$-direction.
%
%
\subsubsection{Interaction picture\label{ip}}
Let us now write the above Hamiltonian in an interaction picture defined by the Hamiltonian $\hat{h}_0=\hat{H}_T+\hbar\tilde\omega_L|e\ra\la e|$, where 
$\tilde\omega_L=(\omega_1+\omega_2)/2$ has been introduced. The interaction Hamiltonian $\hat{H}_I=e^{i\hat{h}_0t/\hbar}(\hat{H}_R-\hat{h}_0)e^{-i\hat{h}_0t/\hbar}$ reads
\begin{eqnarray}
\label{H_int_time_dep}
\hat{H}_I(t)&=&-\hbar\tilde\Delta|e\ra\la e|
\nonumber\\
&+&\sum_{j=1}^2\frac{\hbar\Omega_j}{2}
\left( e^{i\eta_j\left[\hat{a}(t)+\hat{a}^\dag(t)\right]}e^{-i(\omega_j-\tilde\omega_L) t}
e^{-i\phi_j}|e\ra\la g| + H.c.\right),
\end{eqnarray}
where $\tilde\Delta=\tilde\omega_L-\omega_e$, and now $\hat{a}(t)=\hat{a} e^{-i\omega t}$, $\hat{a}^\dag(t)=\hat{a}^\dag e^{i\omega t}$ are the time dependent Heisenberg annihilation and creation operators respectively. Note also that fast oscillating off-resonant $e^{\pm i(\omega_j+\tilde\omega_L)t}$ terms have been neglected in the rotating wave approximation (RWA). The parameter $\eta_j=k_jx_0$ is known as the Lamb-Dicke (LD) parameter,
where $x_0=\sqrt{\hbar/2m\omega}$ is the extension (square root of the variance) of the ion's ground state, i. e., 
$\hat{x}=x_0(\hat{a}+\hat{a}^\dag)$.

%
%
%
\subsubsection{Adiabatic elimination and effective Hamiltonian}
For a general wavefunction (in the corresponding interaction picture) such as
\begin{equation}
|\psi_I(t)\ra=\sum_{n=0}^\infty \left[g_n(t)|g,n\ra+e_n(t)|e,n\ra\right] 
\end{equation}
the differential equations of motion for the probability amplitudes $g_n(t)$ and $e_n(t)$ are obtained from the Schr\"odinger equation $i\hbar\partial_t|\psi_I(t)\ra=\hat{H}_I|\psi_I(t)\ra$,
\begin{eqnarray}
\label{system_gn}
i\dot g_n(t)&=&\frac{1}{2}\sum_{j=1}^2\sum_{n'=0}^\infty
\Omega_je^{i(\theta_jt+\phi_j)}\la n|e^{-i\eta_j\left[\hat{a}(t)+\hat{a}^\dag(t)\right]}|n'\ra e_{n'}(t),\\
i\dot e_n(t)&=&-\tilde\Delta e_n(t)+\frac{1}{2}\sum_{j=1}^2\sum_{n'=0}^\infty
\Omega_je^{-i(\theta_jt+\phi_j)}\la n|e^{i\eta_j\left[\hat{a}(t)+\hat{a}^\dag(t)\right]}|n'\ra g_{n'}(t),
\label{system_en}
\end{eqnarray}
where $\theta_j=\omega_j-\tilde\omega_L$. For large detunings, i. e., for $|\tilde\Delta|\gg\Omega_j,\omega$, see Fig. \ref{level_scheme_fig}, and for an ion initially in the ground state one may assume that the excited state $|e\ra$ is scarcely populated and it may be adiabatically eliminated. 
Then, setting $\dot e(t)=0$, $e_n(t)$ may be written as a function of the $g_{n'}(t)$ from Eq. (\ref{system_en}), and substituting this result into  (\ref{system_gn}) there results a differential equation for the ground state probability amplitude,
\begin{equation}
%
i\dot g_n(t)=s g_n(t)+\frac{\tilde\Omega}{2}\sum_{n'=0}^\infty \mathcal{F}_{n,n'}(t)g_{n'}(t),
\end{equation}
where 
\begin{eqnarray}
s&=&\frac{\Omega_1^2+\Omega_2^2}{4\tilde\Delta},
\\
\mathcal{F}_{n,n'}(t)&=&\la n|e^{-i\tilde\eta\left[\hat{a}(t)+\hat{a}^\dag(t)\right]}|n'\ra e^{i(\tilde\delta t+\tilde\phi)}
+\la n|e^{i\tilde\eta\left[\hat{a}(t)+\hat{a}^\dag(t)\right]}|n'\ra e^{-i(\tilde\delta t+\tilde\phi)},
\end{eqnarray}
and where the effective two-photon Raman parameters, denoted by tildes, 
are given by
\begin{eqnarray}
\tilde\delta&=&\omega_1-\omega_2,
\nonumber\\
\tilde\eta&=&\eta_1-\eta_2,
\nonumber\\
\tilde\phi&=&\phi_1-\phi_2,
\nonumber\\
\frac{\tilde\Omega}{2}&=&\frac{\Omega_1\Omega_2}{4\tilde\Delta}.
\end{eqnarray}
The equation for the ground state probability amplitude corresponds to an effective Hamiltonian 
\begin{eqnarray}
\label{effective_ham_g}
\hat{H}_{eff}&=&\hbar s|g\ra\la g|+
\frac{\hbar\tilde\Omega}{2}\left(e^{i\tilde\eta[\hat{a}(t)+\hat{a}^\dag(t)]}e^{-i(\tilde\delta t+\tilde\phi)}+H.c\right)|g\ra\la g|.
\end{eqnarray}
Note that the Stark-Shift produced by off resonant driving is included in $s$, which is a constant of motion and produces no effect on the Raman coupling between sidebands. We have thus adiabatically eliminated the excited state $|e\ra$ ending with a Hamiltonian of the same form as (\ref{H_int_time_dep}) where the transitions between electronic levels are not present.
%
%
\subsubsection{Two-photon Jaynes-Cummings Hamiltonian in the Raman Scheme: Vibrational RWA}
Using the Baker-Campbell-Hausdorff (BCH) identity, the exponential in the effective Hamiltonian (\ref{effective_ham_g}) may be expanded in power series
of $\tilde\eta$ \cite{Orszag,LME08},
\begin{eqnarray}
\label{time_dep_ham_02}
\hat{H}_{eff}=\frac{\hbar\tilde\Omega}{2}
\left[e^{-\tilde\eta^2/2}\sum_{nn'}
\frac{\left(i \tilde\eta\right)^{n+n'}}{n!n'!}  \hat{a}^{\dag n}\! \hat{a}^{n'}
e^{i (n-n') \omega t}e^{-i\tilde\delta t}e^{-i\tilde\phi}+H.c\right].
\end{eqnarray}
If the effective detuning is $\tilde\delta=\omega_1-\omega_2=2\omega$, the second blue sideband becomes resonant, and we may neglect rapidly oscillating terms in a second or vibrational RWA \cite{LME08}. The above Hamiltonian is then simplified to a two-photon Jaynes-Cummings-like Hamiltonian without electronic transitions. To leading order in $\tilde\eta$ it takes the form 
\begin{equation}
\hat{H}_{2B}=\tilde\eta^2\frac{\hbar\tilde\Omega}{4}\left(\hat{a}^{\dag 2} e^{i\tilde\phi}+\hat{a}^2 e^{-i\tilde\phi}\right)
=i\hbar\frac{\tilde\eta^2\tilde\Omega}{4}\left(\hat{a}^2 -\hat{a}^{\dag 2}\right),
\label{2b}
\end{equation}
where for the last step a relative phase between the applied fields  $\tilde\phi=\phi_1-\phi_2=-\pi/2$ has been assumed.
%
\subsubsection{Validity for time-dependent $\omega$}
Unfortunately the above formal manipulations and approximations 
cannot be carried out in general for a time dependent $\omega$. 
The interaction picture performed in \ref{ip}, in particular, assumes 
a constant $\hat{h}_0$. A time dependent one would require a more complex 
approach with time-ordering operators \cite{Glauber}.
Similarly, the vibrational rotating  
wave approximation requires the stability of the frequency for times larger than a period to avoid off-resonant couplings. 
One may still obtain (\ref{2b}) for a sufficiently slowly varying 
$\omega$, the criterion being   
that the change of the time-dependent trapping frequency in one time period $T$ has to be much smaller than the frequency itself. We can write this condition as $\dot\omega(t) T\ll\omega(t)$ or 
\begin{equation}
\frac{\dot\omega(t)}{\omega(t)^2}\ll1,
\end{equation}
which turns out to be the adiabaticity condition for the harmonic oscillator. 
Of course, if satisfied, the whole enterprise of applying the TT method would be useless. These arguments are far from constituting a proof that the TT method cannot be implemented for the harmonic oscillator. They simply leave this as an open 
question. 
%
%
%
%
%
%
%
%
%
\section{Engineering the Lewis-Riesenfeld invariant}
In this section we describe a different method for transitionless 
dynamics of the harmonic oscillator \cite{harmo}.   
A harmonic oscillator such as $H_0(t)$ in Eq. (\ref{ho}) has 
the following time dependent invariant \cite{LR69}
\beq
I(t)=\frac{1}{2}\left(\frac{\hat{x}^2}{b^2} m \omega_0^2+\frac{1}{m}
\hat{\pi}^2\right),
\eeq
where $\hat{\pi}=b(t)\hat{p}-m\dot{b}\hat{x}$ plays the role of a momentum conjugate to $\hat{x}/b$, the dots are derivatives with respect to time, and $\omega_0$ is in principle an arbitrary constant. The scaling, dimensionless function $b=b(t)$  satisfies the subsidiary condition 
\beq\label{subsi}
\ddot{b}+\omega(t)^2 {b}=\omega_0^2/b^3,
\eeq
an Ermakov equation where real solutions must be chosen to make $I$ Hermitian. 
$\omega_0$ is frequently rescaled to unity by a scale transformation of $b$ \cite{LR69}.
Other common and convenient choice, which we shall adopt here,    
is $\omega_0=\omega(0)$.  
The eigenstates of $I(t)$ become, with appropriate phase factors, solutions of the 
time-dependent Schr\"odinger equation,   
%
\beqa
\Psi_n (t,x) &=& \left(\frac{m\omega_0}{\pi\hbar}\right)^{1/4} 
\!\frac{1}{(2^n n! b)^{1/2}}
\fexp{-i (n+1/2) \int_0^t dt'\, \frac{\omega_0}{b(t')^2}}
\\
&\times&\fexp{i \frac{m}{2\hbar}\left(\frac{\dot{b}}{b(t)}
+
\frac{i\omega_0}{b^2}\right)x^2}
H_n\left[\left(\frac{m\omega_0}{\hbar}\right)^{1/2}\frac{x}{b}\right], 
\label{emode}
\eeqa
%
and form a complete basis to expand any time-dependent state, 
$\psi(x,t)=\sum_n c_n \Psi_n(x,t)$, with the amplitudes $c_n$ constant.  
A method to achieve frictionless, population preserving processes is to leave 
$\omega(t)$ undetermined first, and then set $b$ so that $I(0)=H_0(0)$ and 
$[I(t_f),H_0(t_f)]=0$. This guarantees that the eigenstates of $I$ and $H_0$ are common at initial and finite times. We can do this by setting 
\beqa
b(0)=1, \dot{b}(0)=1, \ddot{b}=0
\nonumber\\
b(t_f)=\gamma=[\omega_0/\omega_f]^{1/2},
\dot{b}(t_f)=0, \ddot{b}(t_f)=0,
\label{bes}
\eeqa
and interpolating $b(t)$ with some real function that satisfies 
these boundary condition. The simplest choice is a polynomial, 
\beq\label{ans}
b(t)=\sum_{j=0}^5 a_j t^j. 
\eeq
Once the $a_j$ are determined from (\ref{bes}),
$\omega(t)$ is obtained from the Ermakov equation (\ref{subsi}),  
and one gets directly a transitionless Hamiltonian
$H_{II}(t)=H_0(t)$ with a local, ordinary, harmonic potential, but note that  
$\omega(t)^2$ may become negative for some time interval, making the
potential an expulsive parabola \cite{harmo,Salomon}.
The II method is thus clearly distinct from from TT and implements a different 
Hamiltonian. Note also,  
by comparison of the coefficients, that the invariant operator $I$ corresponding to $H_{II}$ is different from 
$H_{TT}$, although they are both generalized harmonic oscillators.  
\subsection{Physical realization}
The TT method only requires the time variation of a parabolic potential. 
Effective harmonic optical traps for neutral atoms may be formed by magnetic and/or optical means and their frequencies are routinely varied in time as part of  
many cold atom experiments. 
In magnetic traps, for example the 
frequency is modulated harmonically to look for collective excitation modes of a condensate \cite{Cornell96}, and ramped down 
adiabatically to change its conditions (critical temperature, particle number, spatial extension) \cite{Ketadiab,Cornell96}, 
or as a preliminary step to superimpose an optical lattice \cite{Cas}. 
Some experiments involve both time-dependent magnetic and optical traps 
or antitraps \cite{ch}. Purely optical traps are also manipulated in time,  
e.g. for adiabatic cooling of single neutral atoms \cite{acoo}. 
In particular laser beams detuned with
respect to the atomic transition form effective potentials for the ground state 
depending on Rabi frequency $\Omega$ and detuning $\Delta$ as $\Omega^2/4\Delta$ by adiabatic elimination of the excited state, thus forming attractive or repulsive 
potentials.  
This effective interaction can be made time dependent by varying the laser intensity, the frequency, or both \cite{Bize},        
since the optical frequencies are many orders of magnitude larger than Rabi frequencies or detunings, and the changes will be slowly varying in the scale of  
optical periods.       
The intensity of a dipole trap can be changed by three or four orders
of magnitude  in $100$ ns using acousto-optics or electro-optics modulators.
To monitor the sign of the square frequencies, one can
superimpose two dipole beams locked respectively on the blue and red
side of the line. By controlling their relative
intensity, one can shape the square
frequencies and their signs at will. 
%
%
%
%
%
%
%
%
%
%
%
%
\section{Discussion}
We have compared and distinguished two different methods: a ``transitionless-tracking'' (TT) algorithm, and an ``inverse-invariant'' (II) method, to achieve transitionless 
dynamics for a fast frequency change of a quantum harmonic oscillator.
They imply different driving Hamiltonians. The one in the II method 
can be implemented for ultracold 
atoms or ions in the laboratory by varying the trap frequency in time along a certain trajectory, and a generalization to Bose Einstein condensates has been worked out \cite{becs}, 
but its extension to other potentials or
systems may be difficult and remains an open question. By contrast, 
we have found some difficulties to realize the TT Hamiltonian for the harmonic oscillator, but the TT method has the advantage of being, at least formally, more generally applicable. The feasibility of the actual realization is quite another matter and has to be studied in each case. An example of application is 
provided in \cite{Berry09} for two-level systems.         
\ack{We acknowledge funding by Projects No. GIU07/40, FIS2009-12773-C02-01,
60806041, 08QA14030, 2007CG52, S30105, 
and Juan de la Cierva Program.}

\appendix
\section{Relation to the squeezing operator}
The evolution operator takes a particularly simple form when using the simplified case 
$E_n(t)=0$, so that $\hat{H}(t)=\hat{H}_1(t)$. Taking into account that $[\hat{H}_1(t),\hat{H}_1(t')]=0$
we can write
\beq
\hat{U}(t)=e^{-i\int_0^t \hat{H}_1(t)dt/\hbar}.
\eeq
This may be evaluated explicitly with (\ref{h1a}) fixing the time of the creation and annihilation 
operators to 0, 
\beq
\hat{U}(t)=e^{\frac{1}{2}\ln\left(\sqrt{\frac{\omega(t)}{\omega(0)}}\right)
\left[a_0^2-(a_0^\dagger)^2\right]}=\hat{S}[r(t)],
\eeq
which is a 
sqeezing operator with real argument $r(t)=\ln\left(\sqrt{\frac{\omega(t)}{\omega(0)}}\right)$.
It is unitary with inverse $[\hat{S}(r)]^{-1}=\hat{S}(-r)$. 
Using the relations 
\beqa
\hat{a}_t^\dagger+\hat{a}_t&=&\sqrt{\frac{\omega(t)}{\omega(0)}}(\hat{a}_0^\dagger+\hat{a}_0)
\nonumber
\\
\hat{a}_t^\dagger-\hat{a}_t&=&\sqrt{\frac{\omega(0)}{\omega(t)}}(\hat{a}_0^\dagger-\hat{a}_0)
\eeqa
and the formal 
properties of $\hat{S}$, see e.g. \cite{bar}, it is easy to prove that 
\beqa
\hat{S}(r)\hat{a}_0\hat{S}(-r)&=&\hat{a}_t,
\nonumber
\\
\hat{S}(r)\hat{a}^\dagger_0\hat{S}(-r)&=&\hat{a}_t.
\eeqa
In fact any combination of powers of $\hat{a}_0$ and $\hat{a}^\dagger_0$ is mapped to the same combination of powers of $\hat{a}_t$ and $\hat{a}^\dagger_t$ by this unitary transformation.   
To show that $|0_t\ra\equiv \hat{S}|0_0\ra$ is indeed the vacuum at time $t$, 
note that 
\beq
\hat{a}_t|0_t\ra=\hat{S}(r)\hat{S}(-r)a\hat{S}(r)|0_0\ra=\hat{S}(r)\hat{a}_0|0_0\ra=0.
\eeq
Similarly we note that, consistently, 
\beqa
\hat{S}(r)|n(0)\ra&=&\hat{S}(r)\frac{1}{\sqrt{n!}}(\hat{a}_0^\dagger)^n|0_0\ra
=\frac{1}{\sqrt{n!}}\hat{S}(r)(\hat{a}_0^\dagger)^n\hat{S}(-r)\hat{S}(r)|0_0\ra
\nonumber
\\
&=&\frac{1}{\sqrt{n!}}(\hat{a}_t^\dagger)^n|0_t\ra=|n(t)\ra.
\eeqa

\section*{References}

\end{document}